\documentstyle{mn}
\input{psfig}
\newcommand{\etal}{{et al.~}}

\newcommand{\kmsmpc}{\>{\rm km}\,{\rm s}^{-1}\,{\rm Mpc}^{-1}}

\newcommand{\Mpc}{\>{\rm Mpc}}

\newcommand{\Msun}{\>{\rm M_{\odot}}}

\newcommand{\beq}{\begin{equation}}
\newcommand{\eeq}{\end{equation}}

\newcommand{\calA}{{^{0.1}{\rm A}}}
\newcommand{\calB}{{^{0.0}{\rm A}}}
\newcommand{\calC}{{\cal C}}

\newcommand{\Ms}{M_*}
\newcommand{\Mh}{M_{\rm h}}
\newcommand{\grone}{\>{^{0.1}(g-r)}}
\newcommand{\grzero}{\>{^{0.0}(g-r)}}

\def\gtsima{$\; \buildrel > \over \sim \;$}
\def\ltsima{$\; \buildrel < \over \sim \;$}
\def\prosima{$\; \buildrel \propto \over \sim \;$}
\def\gsim{\lower.7ex\hbox{\gtsima}}
\def\lsim{\lower.7ex\hbox{\ltsima}}
\def\simgt{\lower.7ex\hbox{\gtsima}}
\def\simlt{\lower.7ex\hbox{\ltsima}}
\def\simpr{\lower.7ex\hbox{\prosima}}
\def\la{\lsim}
\def\ga{\gsim}
\def\lta{\la}
\def\gta{\ga}

\newcommand{\apj}{ApJ}
\newcommand{\apjs}{ApJS}
\newcommand{\aj}{AJ}
\newcommand{\mnras}{MNRAS}
\newcommand{\aap}{A\&A}

\newdimen\hssize
\newdimen\hdsize
\newcommand{\Mhsat}{M_{\rm h,sat}}
\newcommand{\Mhcen}{M_{\rm h,cen}}
\begin{document}


\title[The Importance of Satellite Quenching] 
      {The Importance of Satellite Quenching for the Build-Up of 
       the Red Sequence of Present Day Galaxies}
\author[van den Bosch et al.]
       {\parbox[t]{\textwidth}{
        Frank C. van den Bosch$^{1}$\thanks{E-mail: vdbosch@mpia.de}, 
        Daniel Aquino$^{1}$,
        Xiaohu Yang$^{2}$, 
        H.J. Mo$^{3}$,\\
        Anna Pasquali$^{1}$,
        Daniel H. McIntosh$^{3}$,
        Simone M. Weinmann$^{4}$,
        Xi Kang$^{1}$}\\
        \vspace*{3pt} \\
       $^1$Max-Planck-Institute for Astronomy, K\"onigstuhl 17, D-69117
           Heidelberg, Germany\\
       $^2$Shanghai Astronomical Observatory; the Partner Group of MPA,
           Nandan Road 80,  Shanghai 200030, China\\
       $^3$Department of Astronomy, University of Massachusetts,
           Amherst MA 01003-9305, USA\\
       $^4$Institute for Theoretical Physics, University of Zurich,
         CH-8057, Zurich, Switzerland}


\date{}
\pagerange{\pageref{firstpage}--\pageref{lastpage}}
\pubyear{2007}

\maketitle

\label{firstpage}


\begin{abstract}
  According to  the current paradigm, galaxies initially  form as disk
  galaxies at  the centers  of their own  dark matter  haloes.  During
  their subsequent  evolution they may  undergo a transformation  to a
  red, early-type galaxy, thus giving  rise to the build-up of the red
  sequence.   Two  important,  outstanding  questions  are  (i)  which
  transformation  mechanisms  are most  important,  and  (ii) in  what
  environment do  they occur.   In this paper  we study the  impact of
  transformation mechanisms  that operate only  on satellite galaxies,
  such as strangulation, ram-pressure stripping and galaxy harassment.
  Using  a large  galaxy group  catalogue constructed  from  the Sloan
  Digital  Sky Survey,  we compare  the colors  and  concentrations of
  satellites galaxies to those of central galaxies of the same stellar
  mass, adopting the hypothesis that the latter are the progenitors of
  the  former.  On  average, satellite  galaxies are  redder  and more
  concentrated  than  central  galaxies  of  the  same  stellar  mass,
  indicating  that  satellite  specific  transformation  processes  do
  indeed operate.   Central-satellite pairs  that are matched  in both
  stellar  mass  and color,  however,  show  no average  concentration
  difference,  indicating that  the  transformation mechanisms  affect
  color  more  than morphology.   We  also  find  that the  color  and
  concentration  differences of  matched  central-satellite pairs  are
  completely  independent of the  halo mass  of the  satellite galaxy,
  indicating  that  satellite-specific  transformation mechanisms  are
  equally  efficient in  haloes of  all masses.   This  strongly rules
  against mechanisms that are thought  to operate only in very massive
  haloes, such  as ram-pressure stripping or  harassment.  Instead, we
  argue that  strangulation is  the main transformation  mechanism for
  satellite galaxies.   Finally, we determine  the relative importance
  of satellite  quenching for  the build-up of  the red  sequence.  We
  find that roughly 70 percent of red sequence satellite galaxies with
  $\Ms  \sim 10^9 h^{-2}\Msun$  had their  star formation  quenched as
  satellites.   This  drops  rapidly  with  increasing  stellar  mass,
  reaching  virtually   zero  at  $\Ms  \sim   10^{11}  h^{-2}\Msun$.  
  Therefore,  a very  significant fraction  of red  satellite galaxies
  were already quenched before they became a satellite.
\end{abstract}




\section{Introduction}
\label{sec:intro}

The  local  population of  galaxies  consists  roughly  of two  types:
early-types,  which have a  spheroidal morphology,  are red,  and have
little or no ongoing star formation, and late-types, which have a disk
morphology, are blue and have  active, ongoing star formation.  In the
current paradigm  of galaxy formation,  it is believed  that virtually
all galaxies  initially form  as late-type, disk  galaxies due  to the
cooling  of gas  with  non-zero angular  momentum  in virialized  dark
matter  haloes.  During the  subsequent hierarchical  evolution, these
late-type galaxies are then transformed into early-types via a variety
of   mechanisms  that   cause  morphological   transformations  and/or
star formation quenching.

Recent studies have shown that the bimodality of the galaxy population
already exists at  least out to $z  \simeq 1$ (e.g.,  Bell \etal 2004;
Tanaka  \etal 2005; Cucciati  \etal  2006; Cooper \etal  2006; Willmer
\etal 2006),  and  that the  total   stellar mass density  on  the red
sequence  has roughly doubled over the  last 6-8 Gyr (e.g., Bell \etal
2004;   Borch \etal 2006;  Zucca \etal  2006; Faber  \etal 2007; Brown
\etal 2007).    This  strongly supports   the  paradigm that  galaxies
initially form as   disks    and are subsequently  transformed    into
early-types.    What is still  largely   unknown, however, is what the
dominant mechanisms are that cause the late- to early-type transition,
and in what kind of environments they mainly operate.
\begin{figure*}
\centerline{\psfig{figure=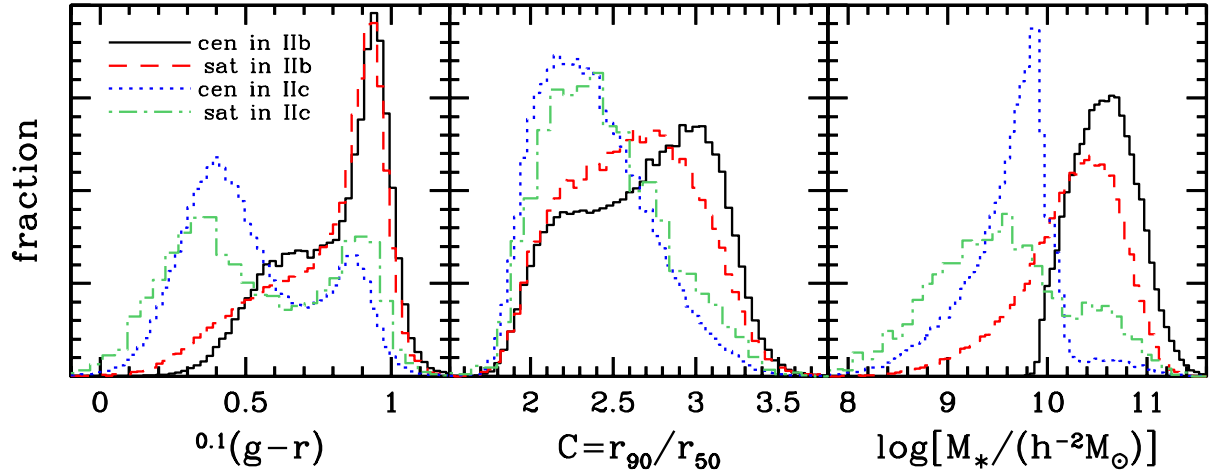,width=0.9\hdsize}}
\caption{The distributions of color  (left panel), concentration 
  (middle  panel)  and stellar  mass  (right  panel)  of centrals  and
  satellites  in samples~IIb and~IIc,  as indicated.   Each individual
  distribution is  normalized to unity;  their relative normalizations
  can be inferred from Table~1.}
\label{fig:galprop}
\end{figure*}

A number  of different mechanisms  have been proposed.  First  of all,
disk galaxies  may be transformed  into early-type spheroids  during a
major  merger  with  another  galaxy  (e.g., Toomre  \&  Toomre  1972;
Negroponte \& White 1983; Kauffmann, White \& Guiderdoni 1993; Hopkins
\etal 2006, 2007a,b).   In order to ensure that  the remnant remains a
red, early-type galaxy, the merging scenario needs to be adjoined with
a mechanism that can prevent any subsequent cooling and star formation
following the merger, which would form  a new disk and make the galaxy
blue again.  Currently, feedback from an active galactic nucleus (AGN)
is believed  to be  the most promising  mechanism that  can accomplish
this (e.g.,  Bower 2006; Croton \etal  2006; Kang, Jing  \& Silk 2006;
Hopkins \etal  2006), though the details remain  unclear.  In addition
to  merging  there  are  a  number  of  transformation  and  quenching
mechanisms which  are special in  that they only operate  on satellite
galaxies.  When  a small halo  is accreted by  a larger halo  its hot,
diffuse  gas  may  be  stripped  thus removing  its  fuel  for  future
star-formation (Larson,  Tinsley \& Caldwell 1980;  Balogh, Navarro \&
Morris  2000).  Following  Balogh  \& Morris(2000)  we  refer to  this
process, that results  in a fairly gradual decline  of the satellite's
star formation  rate, as `strangulation'.  When  the external pressure
is sufficiently high, ram-pressure may also remove the cold gas of the
satellite  galaxy (e.g.,  Gunn \&  Gott 1972;  Quilis, Moore  \& Bower
2000; Hester  2006a), resulting in  an extremely fast quenching  of it
star formation.  In  what follows we refer to  the complete removal of
the  cold  gas  reservoir  as  `ram-pressure  stripping'.   These  two
quenching mechanisms  can transit blue  satellite galaxies to  the red
sequence, but they  do not have a (significant)  impact on the overall
morphology  of the  satellite galaxy.   However, there  are additional
satellite-specific mechanisms that  can transform disks into spheroids
(or at  least, result in  more concentrated morphologies).   Along its
orbit a  satellite galaxy is subject  to tidal forces  which may cause
tidal  stripping  and  heating.   In  addition,  the  morphologies  of
satellite galaxies  can be strongly affected by  the cumulative effect
of  many   high  speed   (impulsive)  encounters,  a   process  called
`harassment' (Farouki  \& Shapiro  1981; Moore \etal  1996).  Finally,
although  most  mergers are  believed  to  involve  a central  galaxy,
satellite-satellite  mergers  do  also  occur  (Makino  \&  Hut  1997;
Somerville \& Primack 1999; McIntosh \etal 2007).
\begin{figure*}
\centerline{\psfig{figure=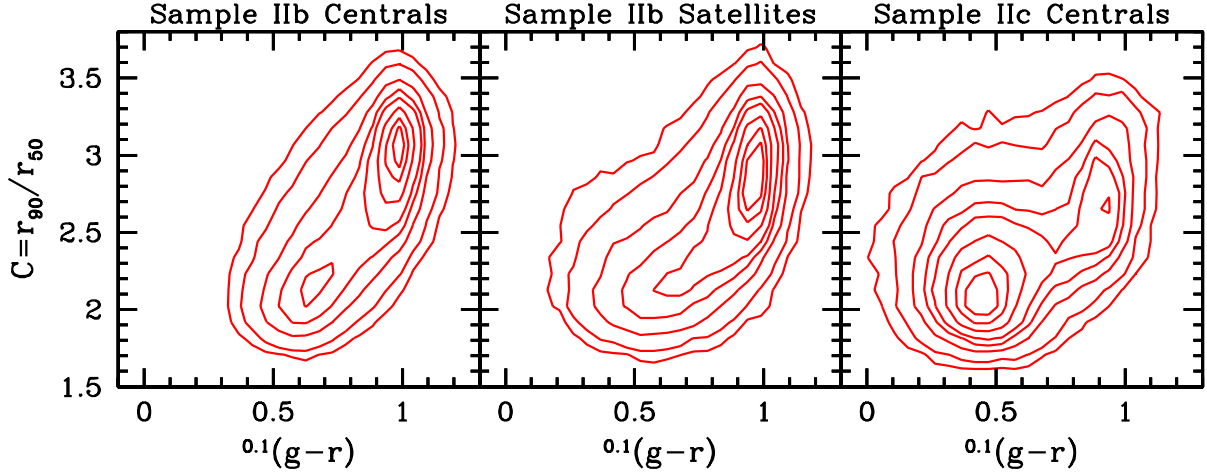,width=0.9\hdsize}}
\caption{The distribution of galaxies in concentration-color   space.  
  Results  are   shown  for  centrals  in   sample~IIb  (left  panel),
  satellites in  sample~IIb (middle panel) and  centrals in sample~IIc
  (right panel).   Note that the  fraction of blue,  low concentration
  galaxies increases  going from  the left panel  to the  right panel,
  which   mainly   reflects   a   trend   in   stellar   mass   (cf.   
  Fig.~\ref{fig:galprop}).}
\label{fig:colcon_cont}
\end{figure*}

The  goals  of  this paper  are  (i)  to  study  the impact  of  these
satellite-specific transformation mechanisms,  (ii) to determine which
process dominates,  (iii) to determine  the fraction of  galaxies that
undergo a  transition from  the blue sequence  to the red  sequence as
satellites, and (iv) to investigate  in what kind of environment these
transitions  take  place.   We  use  a large  galaxy  group  catalogue
constructed from the Sloan Digital  Sky Survey (SDSS; York \etal 2000;
Stoughton  \etal 2002)  by Yang  \etal (2007).   This  group catalogue
yields  information  regarding the  halo  mass  in  which each  galaxy
resides, and  allows us to  split the galaxy population  into centrals
and satellites.  We use this  catalogue to study how the properties of
galaxies (in  particular their  color and concentration)  change after
they are  accreted into a larger  halo.  We compare  the properties of
satellite galaxies with those of  central galaxies of the same stellar
mass, under the hypothesis that  the latter are the progenitors of the
former,  in a  statistical sense.   We also  examine what  fraction of
galaxies  undergoes a  transition from  the blue  sequence to  the red
sequence after having been accreted, by comparing the red fractions of
centrals and satellites of the same stellar mass.  Finally, we use the
group catalogue  to examine how the  transformation efficiency depends
on the  mass of the  halo into which  a galaxy is accreted.   All this
information  combined sheds  important  light on  the efficiency  with
which   the  various   satellite-specific   transformation  mechanisms
discussed above operate.

Throughout  this paper  we adopt  a flat  $\Lambda$CDM  cosmology with
$\Omega_m=0.238$ and  $\Omega_{\Lambda} = 0.762$  (Spergel \etal 2007)
and we express units that depend on the Hubble constant in terms of $h
\equiv H_0/100\kmsmpc$.  In addition, we use `$\log$' as shorthand for
the 10-based logarithm.

\section{Data}
\label{sec:data}

The analysis presented  in this paper is based on  the SDSS DR4 galaxy
group  catalogue of  Yang  \etal (2007;  hereafter  Y07).  This  group
catalogue is constructed applying  the halo-based group finder of Yang
\etal (2005a) to the  New York University Value-Added Galaxy Catalogue
(NYU-VAGC;  see Blanton  \etal  2005a),  which is  based  on SDSS  DR4
(Adelman-McCarthy \etal  2006).  From this catalogue  Y07 selected all
galaxies  in  the Main  Galaxy  Sample  with  an extinction  corrected
apparent magnitude brighter than $m_r=18$, with redshifts in the range
$0.01 \leq  z \leq 0.20$ and  with a redshift  completeness $\calC_z >
0.7$.   This sample  of  galaxies  is used  to  construct three  group
samples: sample I, which only uses the $362356$ galaxies with measured
redshifts from the SDSS, sample II which also includes $7091$ galaxies
with  SDSS  photometry  but  with  redshifts  taken  from  alternative
surveys, and sample III  which includes an additional $38672$ galaxies
that lack a redshift due  to fiber-collisions, but which we assign the
redshift  of  its nearest  neighbor  (cf.   Zehavi  \etal 2002).   The
present  analysis  is  based  on   all  galaxies  in  sample  II  with
$m_r<17.77$, which consists of  $344348$ galaxies, but we have checked
that samples I and III give results that are almost indistinguishable.
\begin{table}
\label{tab:samples}
\caption{Galaxy Samples}
\begin{tabular}{cccccl}
   \hline
ID  & \#gal & \#grp & \#cen & \#sat & description \\
\hline\hline
IIa & 344348 & 281089 & 280879 & 63469 & entire sample \\
IIb & 278085 & 218282 & 218103 & 59982 & with group mass \\
IIc & 66263  & 62807  & 62776  & 3487  & no group mass \\
\hline
\end{tabular}
\medskip

\begin{minipage}{\hssize}
  {\it Notes:}  Properties of  the three galaxy  samples used  in this
  paper. Columns (2)  to (5) list the numbers  of galaxies, of groups,
  of central  galaxies and  of satellite galaxies,  respectively.  The
  sixth column gives a brief description of the sample.  Sample~IIa is
  the entire galaxy sample.  Sample~IIb consists of those galaxies for
  which the group  has an assigned halo mass.   Sample~IIc consists of
  those galaxies  for which the group  does not have  an assigned halo
  mass. Thus sample~IIa is simply the sum of samples~IIb and~IIc.
\end{minipage}

\end{table}

The magnitudes and colors  of all galaxies are  based on  the standard
SDSS Petrosian technique  (Petrosian  1976; Strauss \etal 2002),  have
been corrected for galactic extinction  (Schlegel, Finkbeiner \& Davis
1998), and have been $K$-corrected and evolution corrected to $z=0.1$,
using the method described  in  Blanton \etal (2003).  Stellar  masses
for  all galaxies  are computed using   the relations between  stellar
mass-to-light  ratio and color of  Bell \etal  (2003; see Appendix~A).
The main  galaxy parameters that we use  in this paper are the stellar
mass,     $\Ms$,  the   $\grone$     color,   and  the   concentration
$C=r_{90}/r_{50}$.  Here $r_{90}$   and  $r_{50}$ are the  radii  that
contain   90 and  50     percent  of the    Petrosian   $r$-band flux,
respectively.  As shown by Strateva \etal  (2001), $C$ is a reasonable
proxy  for Hubble type,  with $C>2.6$  corresponding  to an early-type
morphology.
\begin{figure*}
  \centerline{\psfig{figure=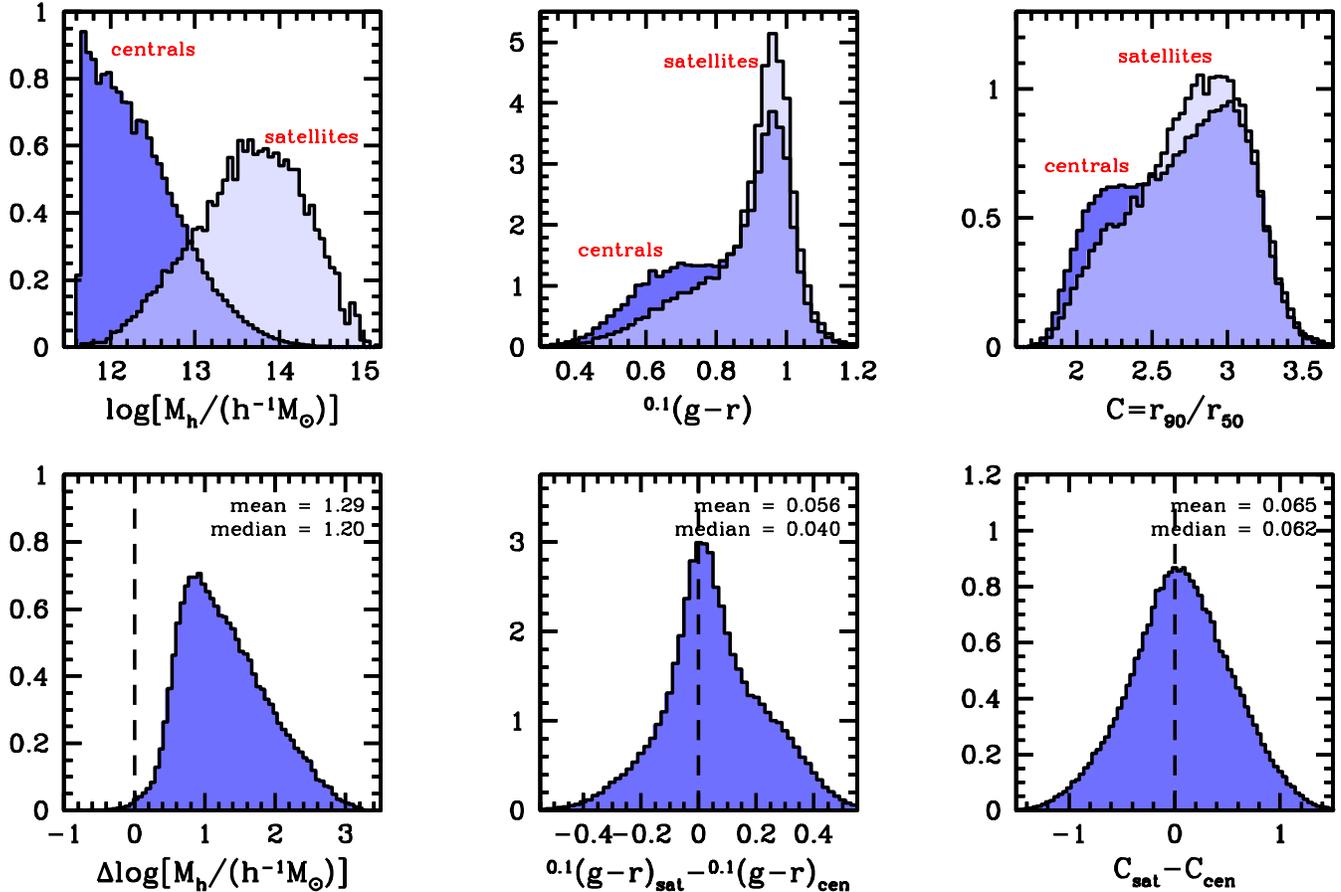,width=\hdsize}}
  \caption{The upper left-hand panel shows the halo mass distributions
    of  the central  and  satellite galaxies  of  sample~IIb that  are
    matched  in  stellar mass.   The  distributions  for centrals  and
    satellites  are  shaded dark  and  light,  respectively, while  an
    intermediate color is used to shade the overlap region.  The lower
    left-hand   panel   shows   the  corresponding   distribution   of
    $\Delta\log\Mh = \Mhsat-\Mhcen$, and shows that satellite galaxies
    live in  haloes that are on  average about one  order of magnitude
    more massive than central galaxies  of the same stellar mass.  The
    middle and right-hand  panels in the upper row  show the color and
    concentration distributions, respectively, of the central galaxies
    (dark shading) and their matched satellites (light shading).  Once
    again,  an intermediate shading  is used  to indicate  the overlap
    regions.  Finally,  the middle and right-hand panels  of the lower
    row  show  the  corresponding   distributions  of  the  color  and
    concentration  differences  between  centrals  and  their  matched
    satellites.   The  mean  and  medians of  both  distributions  are
    indicated.}
\label{fig:diff}
\end{figure*}

As described in Y07, the majority of the  groups in our catalogue have
two estimates of their  dark matter halo mass $\Mh$:  one based on the
ranking of its total characteristic luminosity, and the other based on
the   ranking of its total  characteristic  stellar mass.  As shown in
Y07, both halo masses agree very well with each other, with an average
scatter that  decreases from $\sim 0.1$  dex  at the  low mass  end to
$\sim 0.05$ dex at the massive end.  In  addition, detailed tests with
mock   galaxy redshift catalogues  have  demonstrated that these group
masses are more reliable than  those based on the velocity  dispersion
of the  group members (Yang \etal 2005b;  Weinmann  \etal 2006a; Y07).
In this paper we adopt  the  group masses based   on the stellar  mass
ranking\footnote{We have verified,   though, that none of our  results
  change  significantly if we   adopt the luminosity-rank based masses
  instead.}.  These masses are  available   for a total  of   $218282$
groups in our sample,  which host a  total of $278085$ galaxies.  This
implies that a total of $66263$ galaxies have been assigned to a group
for which no reliable mass estimate is available.

\subsection{Sample Definition and Global Properties}
\label{sec:global}

In  this paper  we consider  three galaxy  samples:  sample~IIa, which
consists of all 344348 galaxies in group sample II with $m_r < 17.77$,
sample~IIb, which only  considers the galaxies in groups  that have an
assigned  mass, and  sample~IIc,  which consists  of  the galaxies  in
groups that  do not  have an assigned  mass.  The properties  of these
three  samples are listed  in Table~1.   In each  sample we  split the
galaxies into  ``centrals'', which are the most  massive group members
in terms  of their stellar  mass, and ``satellites'', which  are those
group members that are not  centrals.  Note that only $5.3$ percent of
the  galaxies in  sample~IIc are  satellites, compared  to  $18.4$ and
$21.6$ percent in samples~IIa and~IIb, respectively.

Fig.~\ref{fig:galprop}    shows    the    (individually    normalized)
distributions of  color, concentration,  and stellar mass  for central
and satellite  galaxies in samples~IIb  and~IIc.  Note that  the color
and concentration distributions of centrals and satellites in the same
sample are  remarkably similar, but  that they are very  different for
different  samples.  In  the  case of  sample~IIb,  both centrals  and
satellites  reveal a  pronounced red  peak  (representing red-sequence
galaxies) and a modest blue tail.  In the case of sample~IIc, however,
the color distribution has a much more pronounced blue peak and a less
significant red-sequence,  while the concentration  distribution peaks
at   significantly  lower   values  (again   for  both   centrals  and
satellites).   As is evident  from the  right-hand panel,  galaxies in
sample~IIc have,  on average, significantly lower  stellar masses than
galaxies  in sample~IIb.  Another  important difference  between these
two samples is the masses of  the haloes in which the galaxies reside. 
More than  90 percent of the  galaxies in sample~IIb  reside in haloes
with $\Mh  \geq 10^{11.8} h^{-1}  \Msun$.  In the case  of sample~IIc,
although  no halo  masses have  been assigned  to these  galaxies, the
majority almost  certainly resides in  low mass haloes with  $\Mh \lta
10^{12}  h^{-1}  \Msun$.   Therefore, Fig.~\ref{fig:galprop}  suggests
that the colors and concentrations  of a galaxy do not depend strongly
on  whether the  galaxy is  a  central or  a satellite,  but are  more
closely related  to their stellar  mass or to  the mass of  their dark
matter halo.

Fig.~\ref{fig:colcon_cont}  shows  the  distribution of  three  galaxy
samples  in color-concentration space.   Results are  shown separately
for  central  galaxies  in  sample~IIb  (left-hand  panel),  satellite
galaxies  in  sample~IIb  (middle   panel)  and  central  galaxies  in
sample~IIc  (right-hand  panel).  There  are clear  differences:  when
moving from the left-hand panel  to the right-hand panel, the fraction
of blue  galaxies with low concentrations  increases profoundly.  This
confirms  the conclusion  reached above:  if the  main  mechanism that
determines whether a galaxy is an early-type (red and concentrated) or
a late  type (blue  and less concentrated)  is related to  whether the
galaxy  is a  central  or a  satellite,  than the  color-concentration
distributions of the left-  and right-hand panels should look similar,
which   is  clearly   not  the   case.    On  the   other  hand,   the
color-concentration    distributions   are   consistent,    at   least
quantitatively,  with  a  picture  in which  the  early-type  fraction
increases monotonically with stellar mass (cf. the right-hand panel of
Fig.~\ref{fig:galprop}).
\begin{figure*}
\centerline{\psfig{figure=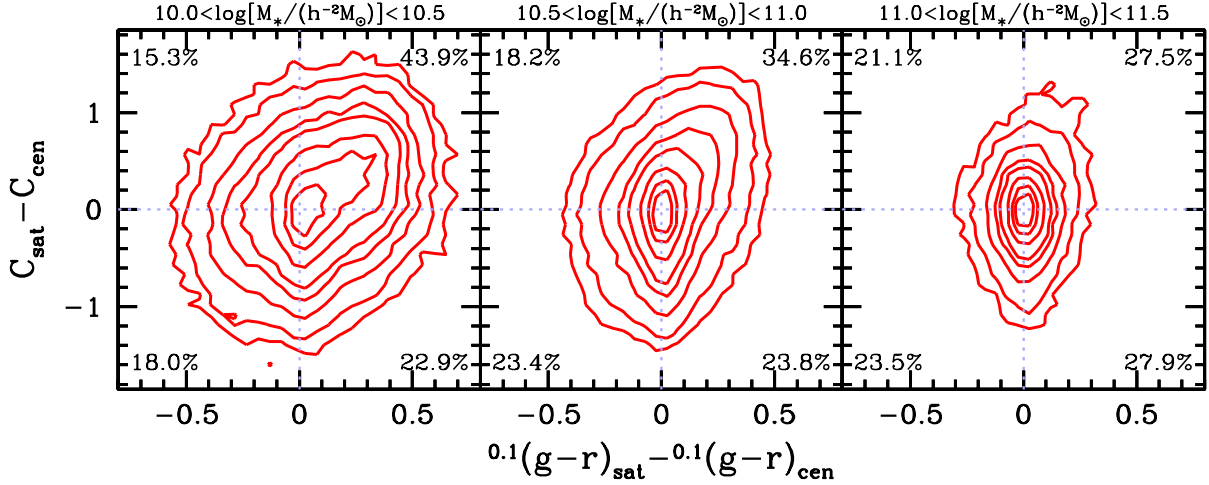,width=0.9\hdsize}}
\caption{Contour plots of  $\grone_{\rm sat}-\grone_{\rm cen}$  versus
  $C_{\rm sat}-C_{\rm cen}$ for three bins  in stellar mass (indicated
  at  the  top   of   each  panel).   The   percentages  of    matched
  central-satellite pairs  in     each of   the  four  quadrants   are
  indicated.}
\label{fig:dcol_dcon}
\end{figure*}

\section{The Impact of Satellite Quenching}
\label{sec:transform}

One of  the main aims of this  paper is to investigate  to what extent
the  colors and concentrations  of central  galaxies change  when they
become a  satellite galaxy (i.e., when  they are accreted  into a more
massive   halo).   To   this  extent   we  compare   the   colors  and
concentrations of  satellite galaxies to those of  central galaxies of
the same  stellar mass, under the  hypothesis that the  latter are the
progenitors  of the  former (in  a  statistical sense).   We start  by
analyzing sample~IIb,  for which we  have halo masses  available.  For
each central galaxy in this sample we randomly pick a satellite galaxy
whose     stellar    mass    matches     that    of     the    central
($\vert\Delta\log\Ms\vert \leq 0.01$)  and we register the differences
in color,  $\grone_{\rm sat}  - \grone_{\rm cen}$,  and concentration,
$C_{\rm  sat}-C_{\rm cen}$, as  well as  the masses  of the  haloes in
which  the  central  and  satellite  reside,  $\Mhcen$  and  $\Mhsat$,
respectively.  The upper  left-hand panel of Fig.~\ref{fig:diff} shows
the halo  mass distributions of  the centrals and satellites  that are
matched in  stellar mass,  while the lower  left-hand panel  shows the
corresponding distribution  of the differences,  $\Delta\log\Mh \equiv
\log\Mhsat - \log\Mhcen$.  Obviously satellite galaxies reside in more
massive  haloes   than  their   central  counterparts,  with   a  mean
$\Delta\log\Mh  $  of  $1.29$.   Note  that  there  are  virtually  no
central-satellite pairs  with $\Delta\log\Mh < 0$.   This is important
since a central galaxy in a halo of mass $\Mh$ is not expected to ever
become a satellite galaxy in a less massive halo.  In other words, the
fact  that  $\Delta\log\Mh  >  0$   for  almost  all  of  our  matched
central-satellite pairs is in line with the idea that centrals are the
progenitors of satellites of the same stellar mass.

The   middle   and   right-hand   panels   in   the   upper   row   of
Fig.~\ref{fig:diff} show the  color and concentration distributions of
the  centrals  and their  matched  satellites.   These are  remarkably
similar,  though the satellites  have a  somewhat more  pronounced red
peak, and  are more skewed towards higher  concentrations.  The middle
panel in  the lower row of Fig.~\ref{fig:diff}  shows the distribution
of the color differences  between our matched central-satellite pairs. 
Although it is clearly skewed towards satellite galaxies being redder,
the mean (median)  of the distribution is only  $0.056$ ($0.040$).  As
is   evident   from    a   comparison   with   Figs.~\ref{fig:galprop}
and~\ref{fig:colcon_cont}, this color  difference is much smaller than
the typical color difference between the red and blue sequences, which
suggests that the fraction of satellite galaxies that undergoes a blue
sequence  to   red  sequence  transition  is   relatively  small  (see
\S\ref{sec:trans}).    Finally,   the   lower  right-hand   panel   of
Fig.~\ref{fig:diff}  shows  the  distribution of  $C_{\rm  sat}-C_{\rm
  cen}$ for our matched central-satellite pairs.  This distribution is
remarkably symmetric with  a mean (median) of only  $0.065$ ($0.062$). 
Once again  this difference is small  compared to the  variance in $C$
shown  in   Figs.~\ref{fig:galprop}  and~\ref{fig:colcon_cont}.   This
suggests that  galaxies do not  undergo a strong change  in morphology
once they become a satellite galaxy.
\begin{figure}
\centerline{\psfig{figure=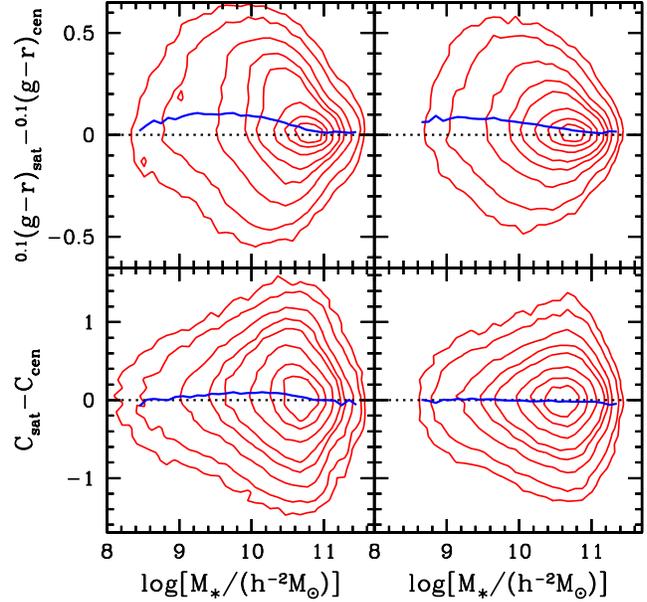,width=\hssize}}
\caption{Contour plots  of $\grone_{\rm sat}-\grone_{\rm  cen}$ (upper
  panels) and $C_{\rm sat}-C_{\rm cen}$ (lower panels) as functions of
  stellar  mass for  central-satellite  pairs in  Sample~IIa.  In  the
  left-hand  panels  centrals  and  satellites have  been  matched  in
  stellar mass only.  In the  right-hand panels they have been matched
  in stellar mass and  concentration (upper-right panel) or in stellar
  mass and color (lower-right  panel).  Blue, solid lines indicate the
  running averages.}
\label{fig:res}
\end{figure}

Fig.~\ref{fig:dcol_dcon} plots  the  distribution of central-satellite
pairs as function  of both the  color difference and the concentration
difference.  Results   are shown for  three  bins in stellar  mass, as
indicated at   the top of  each   panel, while  we also  indicate  the
percentage  of       pairs   in  each   quadrant.       For   $10.0  <
\log[\Ms/(h^{-2}\Msun)] < 10.5$ the distribution is clearly asymmetric
with respect to the   origin;  almost half   ($43.9$ percent) of   all
central-satellite pairs lie    in the first quadrant (satellites   are
redder and more concentrated), compared to  only $18.0$ percent in the
third  quadrant (satellites are   bluer and less concentrated).   This
clearly demonstrates that   low  mass galaxies undergo  a  significant
transformation once they become  a  satellite galaxy.  Also, the  fact
that the fourth quadrant  contains a significantly larger fraction  of
galaxies than the  second quadrant  indicates that the  transformation
mechanism(s)  have  a larger impact  on color  than  on concentration.
Similar trends are   apparent in the  middle   panel, corresponding to
$10.5  <   \log[\Ms/(h^{-2}\Msun)]  <  11.0$,  albeit less pronounced.
Finally, in the most  massive bin (right-hand panel), the distribution
starts to look remarkably  symmetric, with only a  very mild excess of
pairs on the red side. Thus, there is a clear stellar mass dependence,
in that less massive galaxies undergo a more pronounced transformation
when they become a satellite galaxy.

Fig.~\ref{fig:res}  shows  another  rendition  of  this  stellar  mass
dependence.     It   shows    the   distributions    of   $\grone_{\rm
  sat}-\grone_{\rm cen}$ (upper  panels) and $C_{\rm sat}-C_{\rm cen}$
(lower  panels) as  functions of  stellar mass,  with the  blue, solid
lines  indicating the  running  averages.  In  order  to maximize  the
dynamic range in  stellar mass, we here use  Sample~IIa which includes
galaxies  in  groups that  lack  an  assigned  halo mass  (Sample~IIb,
though, yields similar results over the stellar mass range in common).
Note that  centrals with $\Ms  \gta 10^{11} h^{-2} \Msun$  have almost
exactly the  same colors and  concentrations as satellite  galaxies of
the same stellar mass (on average).  Towards lower $\Ms$, however, the
satellites become  redder and more concentrated compared  to a central
galaxy of the  same stellar mass.  At $\Ms  \sim 10^{10} h^{-2} \Msun$
satellites are,  on average, $\sim  0.1$ magnitudes redder and  have a
concentration parameter  that is $\sim 0.1$ larger  than their central
counterparts. Note also that there is  a weak trend that the color and
concentration differences become smaller  again for $\Ms \lta 3 \times
10^9 h^{-2}  \Msun$.  In particular,  centrals and satellites  with $8
\lta \log[\Ms/(h^{-2}\Msun)] \lta 9$ show no significant difference in
concentration.

To  a  good approximation,  ram-pressure  stripping and  strangulation
should  cause a  change  in  color, but  leave  the concentration  and
stellar  mass   largely  intact.    Therefore,  if  either   of  these
transformation  mechanisms  is efficient,  we  would expect  satellite
galaxies to be significantly redder  than central galaxies of the same
stellar  mass {\it  and  the  same concentration}.   To  test this  we
proceed  as  follows.   For  each  central galaxy  in  sample~IIa,  we
randomly pick a satellite  galaxy whose stellar mass and concentration
match that  of the central  ($\vert \Delta\log\Ms\vert \leq  0.01$ and
$\vert\Delta C\vert \leq 0.01$), and we register the color difference.
The upper  right-hand panel of Fig.~\ref{fig:res}  shows the resulting
distribution of the color differences as function of stellar mass. The
running average is slightly smaller than  in the case in which we only
match  the  central-satellite  pairs   in  stellar  mass,  though  the
differences  are   only  modest.    The  lower  right-hand   panel  of
Fig.~\ref{fig:res}   shows  the   differences  in   concentration  for
central-satellite pairs matched in both stellar mass and color ($\vert
\Delta\log\Ms\vert  \leq  0.01$   and  $\vert\Delta\grone  \vert  \leq
0.01$).  Remarkably,  the running average is almost  exactly zero over
the entire range  in stellar masses probed.  This  suggests that there
are  no satellite  galaxies  that undergo  a  change in  concentration
without a change in color.   On the other hand, a significant fraction
of  satellite   galaxies  has  undergone  a  change   in  color  while
maintaining the same  concentration.  This suggests that strangulation
and/or ram-pressure  stripping are more  efficient than transformation
mechanisms that affect morphology,  such as harassment. It also nicely
explains why broad-band colors are more predictive of environment than
morphology, as noted  by Hogg \etal (2003) and  Blanton \etal (2005b),
and is in good agreement with the results of Hester (2006b)

\begin{figure}
\centerline{\psfig{figure=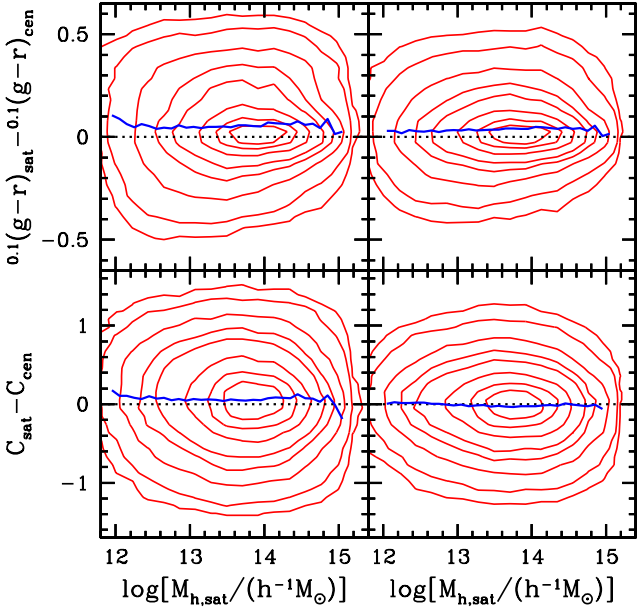,width=\hssize}}
\caption{Contour plots of  $\grone_{\rm sat}-\grone_{\rm cen}$  (upper
  panels)   and   $C_{\rm  sat}-C_{\rm   cen}$   (lower  panels)   for
  central-satellite pairs in sample~IIb  as functions of the halo mass
  of the satellites.  In  the left-hand panels centrals and satellites
  have been  matched in stellar  mass only.  In the  right-hand panels
  they   have  been   matched  in   stellar  mass   and  concentration
  (upper-right  panel)  or  in  stellar mass  and  color  (lower-right
  panel).  Blue, solid lines indicate the running averages.}
\label{fig:delta_cont}
\end{figure}

Having  examined the  stellar  mass  dependence, we  now  turn to  the
dependence on  the halo mass  of the satellite galaxy,  $\Mhsat$.  The
left-hand  panels  of  Fig.~\ref{fig:delta_cont}  show the  color  and
concentration  differences of  central-satellite  pairs in  sample~IIb
that have been  matched in stellar mass as functions  of the halo mass
of  the   satellite  galaxy.   The  running   averages  are  virtually
independent of  $\Mhsat$, but clearly offset from  zero.  This absence
of any dependence on $\Mhsat$  suggests that the efficiency with which
the transformation  mechanisms operate are  independent of halo  mass. 
As we discuss in \S\ref{sec:concl} this puts strong constraints on the
viability of the various transformation mechanisms.

In the  right-hand panels  of Fig.~\ref{fig:delta_cont} we  repeat the
same  exercise, but this  time we  have matched  the central-satellite
pairs in both stellar mass and concentration (upper-right panel) or in
both stellar mass and color (lower-right panel).  Once again, there is
no significant  dependence on $\Mhsat$.  Note also  that, in agreement
with the results shown in the lower-right panel of Fig.~\ref{fig:res},
centrals and satellites with the  same stellar mass and the same color
show no difference in concentration, on average.

\section{Transition Fractions}
\label{sec:trans}

\begin{figure}
\centerline{\psfig{figure=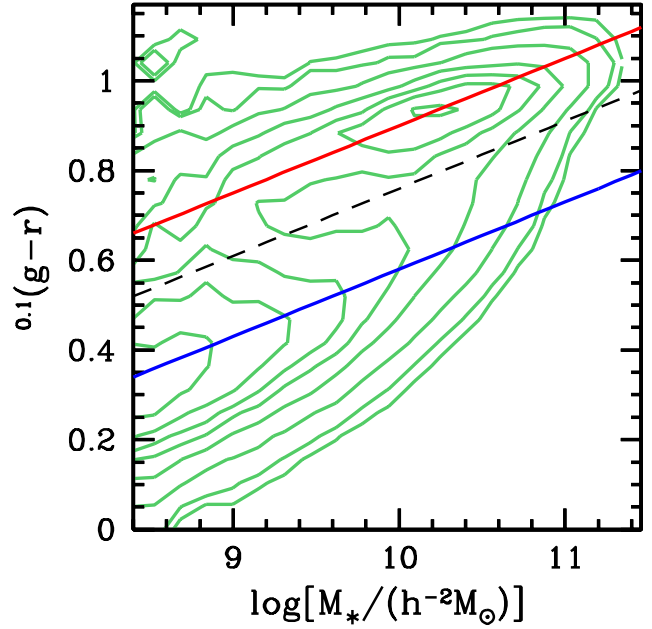,width=\hssize}}
\caption{Contour plot of the $1/V_{\rm max}$-weighted distribution of
  galaxies in sample~IIa  as function of color and  stellar mass.  The
  red and blue lines indicate the red and blue sequences, respectively
  (fit by eye), while the dashed black line indicates our split in red
  (above) and blue (below) galaxies.}
\label{fig:cc_mstar}
\end{figure}

Having determined  the average differences in  color and concentration
between  centrals and  satellites of  the  same stellar  mass, we  now
determine the  {\it fractional} importance of  satellite quenching for
the build-up  of the red  sequence.  We do  this by comparing  the red
fractions of  centrals and satellites. As  a first step,  we split our
galaxy population  in red and  blue galaxies.  Fig.~\ref{fig:cc_mstar}
shows the  color-stellar mass relation  of all galaxies  in sample~IIa
(centrals and satellites combined).  Here we have weighted each galaxy
by $1/V_{\rm max}$, where $V_{\rm  max}$ is the comoving volume of the
Universe out  to a comoving distance  at which the  galaxy would still
have made the selection criteria of our sample.  This weighting scheme
corrects the  sample for  Malmquist bias, and  for the fact  that blue
galaxies can  be probed out to  higher redshifts than  red galaxies of
the  same  stellar  mass  (see Appendix~A).   The  color-stellar  mass
distribution clearly reveals the  bimodality of the galaxy population:
massive galaxies  are mainly red,  while low-mass galaxies  are mainly
blue.  The red and blue lines in Fig.~\ref{fig:cc_mstar} delineate the
red and blue sequences given by
\begin{equation}\label{colcut}
\grone = \calA + 0.15 \left( \log[\Ms/(h^{-2}\Msun)] - 10.0\right)\,,
\end{equation}
with  $\calA=\calA_{\rm red}=0.90$ and  $\calA=\calA_{\rm blue}=0.58$,
respectively.  These relations are simply fit by eye, and merely serve
to illustrate the rough stellar mass dependence of the sequences.  The
dashed line indicates the color-cut adopted in this paper to split the
population  in  red  and   blue  galaxies,  and  is  parameterized  by
eq.~(\ref{colcut}) with  $\calA=\calA_{\rm cut} =  0.76$.  Again, this
cut  is  chosen  somewhat  arbitrarily,  though our  results  are  not
sensitive  to   the  exact  value   of  $\calA_{\rm  cut}$   adopted.  
Alternatively, we could have  fitted the color distribution at various
stellar mass bins with a bi-Gaussian function, and use the bisector of
the medians of the two Gaussians to split the galaxy population in red
and blue  (cf. Baldry  \etal 2004; Li  \etal 2006).  Although  this is
less ambiguous  than our simple fit-by-eye,  it has a  problem in that
the  color-distribution  is no-longer  bimodal  at  $\Ms \gta  10^{11}
h^{-2} \Msun$. Consequently, the bisector basically splits the massive
end of the red sequence in  two.  We therefore believe that our method
for splitting the galaxy population in red and blue is more sensible.

Using the above color-cut and  our group catalogue, we now investigate
the importance of  satellite-specific transformation processes for the
build-up of the red sequence.   In what follows, we use the subscripts
`r' and  `b' to refer to `red'  and `blue', and the  subscripts `c' and
`s'  to  refer  to  `centrals' and  `satellites',  respectively.   All
fractions  are properly  weighted by  $1/V_{\rm max}$  to  correct for
Malmquist bias.  For example, the  fraction of satellites that are red
is given by
\begin{equation}\label{frs}
f_{\rm r|s}(\Ms) = \sum\limits_{i=1}^{N_{\rm rs}} w_i /  
                   \sum\limits_{i=1}^{N_{\rm s}} w_i\,,
\end{equation} 
where  $N_{\rm rs}$ is  the number  of red  satellites of  mass $\Ms$,
$N_{\rm s}$  is the corresponding  number of satellites, and  $w_i$ is
the $1/V_{\rm  max}$ weight  of galaxy $i$.   Note that  $f_{\rm r|s}$
should not be confused with  $f_{\rm rs} = \sum_{i=1}^{N_{\rm rs}} w_i
/ \sum_{i=1}^{N_{\rm  tot}} w_i$, which  is the fraction of  {\it all}
galaxies of mass $\Ms$ that are red satellites, or with $f_{\rm s|r} =
\sum_{i=1}^{N_{\rm rs}}  w_i /  \sum_{i=1}^{N_{\rm r}} w_i$,  which is
the satellite  fraction of  red galaxies (with  $N_{\rm r}$  the total
number of  red galaxies of  mass $\Ms$).  Errors are  determined using
the jackknife  technique.  We divide  the group catalogue  into $N=20$
subsamples  of  roughly  equal   size,  and  recalculate  the  various
fractions 20  times, each time leaving  out one of the  20 subsamples. 
The jackknife estimate of the standard deviation then follows from
\begin{equation}
\label{jack}
\sigma_f = \sqrt{{N-1 \over N} \sum_{i=1}^N \left(f_i - \bar{f}\right)^2}
\end{equation}
with  $f_i$  the fraction  obtained  from  jackknife  sample $i$,  and
$\bar{f}$ is the average.

\begin{figure*}
\centerline{\psfig{figure=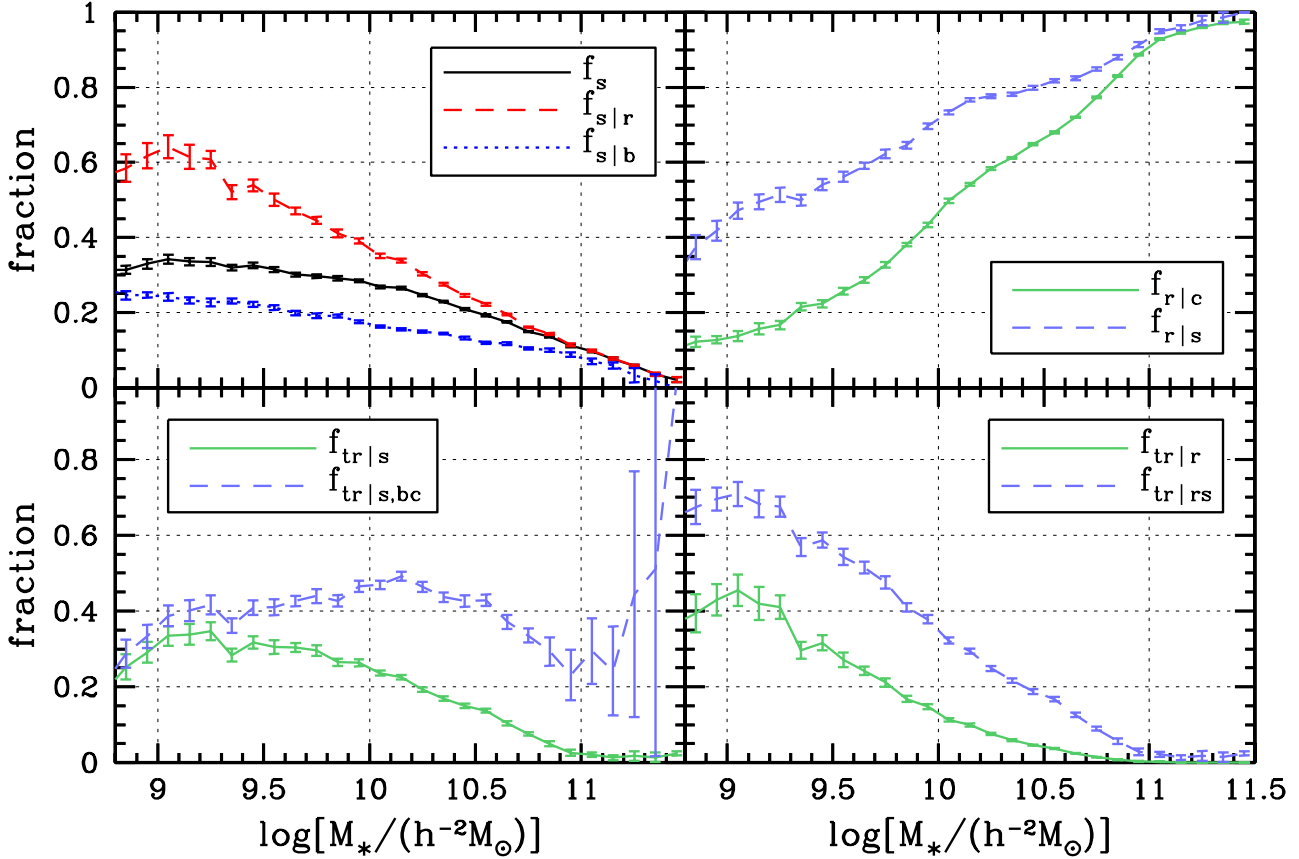,width=0.96\hdsize}}
\caption{{\it Upper left:} The satellite fractions of all galaxies
  ($f_{\rm  s}$, solid black  line), of  red galaxies  ($f_{\rm s|r}$,
  dashed red  line) and of  blue galaxies ($f_{\rm s|b}$,  dotted blue
  line)  as functions  of stellar  mass.  {\it  Upper right:}  The red
  fractions  of  centrals  ($f_{\rm   r|c}$,  solid  green  line)  and
  satellites  ($f_{\rm  r|s}$,  dashed  blue  line).   Note  that  the
  fraction  of red  satellites is  higher than  that of  red centrals,
  indicating that a certain fraction of galaxies has to transit to the
  red sequence once they become a satellite galaxy.  {\it Lower left:}
  the transition  fraction of satellites ($f_{\rm  tr|s}$, solid green
  line),  and the transition  fraction of  satellites that  were still
  blue at the time of accretion ($f_{\rm tr|s,bc}$, solid green line).
  {\it Lower right:} the  transition fraction of red sequence galaxies
  ($f_{\rm tr|r}$,  solid green line)  and of red  sequence satellites
  ($f_{\rm tr|rs}$, dashed blue line).  In all cases the errorbars are
  obtained using  20 jackknife  samples.  All fractions  are $1/V_{\rm
    max}$ weighted, and are listed in Table~2.}
\label{fig:transit}
\end{figure*}

The black line in  the upper left-hand panel of Fig.~\ref{fig:transit}
shows  the satellite  fraction, $f_{\rm  s}$, as  function  of stellar
mass.  This decreases from $\sim 0.35$ at $\Ms = 10^9 h^{-2} \Msun$ to
virtually zero at  $\Ms = 3 \times 10^{11}  h^{-2} \Msun$.  The dashed
(red)  and dotted  (blue) lines  show the  satellite fractions  of red
($f_{\rm  s|r}$)  and blue  ($f_{\rm  s|b}$)  galaxies, respectively.  
Clearly, the  satellite fraction is  higher for red galaxies  than for
blue  galaxies.   Note that  these  satellite  fractions  are in  good
agreement with  results obtained from galaxy  clustering (Cooray 2006;
Tinker \etal  2007; van den  Bosch \etal 2007) and  from galaxy-galaxy
lensing (Mandelbaum \etal 2006).

The  upper right-hand  panel of Fig.~\ref{fig:transit}   plots the red
fractions of   central  galaxies  ($f_{\rm  r|c}$,  solid   line)  and
satellites  ($f_{\rm  r|s}$,  dashed line).  In   both  cases  the red
fractions increase strongly   with  increasing stellar    mass.  Above
$10^{11}  h^{-2} \Msun$ the red   fractions of centrals and satellites
are indistinguishable and close to unity. However, at the low mass end
the red fractions are clearly  higher for satellites than for centrals
(see   also   Weinmann \etal   2006a).     Under the  assumption  that
present-day central   galaxies   are the  progenitors  of  present-day
satellite galaxies of  the  same stellar mass,   this  implies that  a
certain fraction of the  satellite galaxies must have transited  from
the blue sequence to the red sequence as satellites.  We now compute a
number  of  different transition  fractions.  First  we  determine the
fraction of all satellite galaxies that have undergone a blue sequence
to red  sequence transition after  their  accretion (i.e., after they
became a satellite galaxy). This fraction is given by
\begin{equation}\label{ftrsat}
f_{\rm tr|s}(\Ms) = f_{\rm r|s}(\Ms) - f_{\rm r|c}(\Ms)
\end{equation} 
and  is  shown  as a  solid  line  in  the  lower left-hand  panel  of
Fig.~\ref{fig:transit}.   Note that  $ f_{\rm  tr|s}(\Ms)$  drops from
$\sim 35$ percent  at $\Ms \simeq 10^9 h^{-2}  \Msun$ to virtually zero
for  $\Ms  \gta  10^{11}   h^{-2}  \Msun$.   However,  this  does  not
necessarily  mean  that   the  transformation  process(es)  work  less
efficiently on  more massive galaxies.  After all,  only galaxies that
are blue  prior to  becoming a satellite  can undergo  the blue-to-red
transition. Therefore, the decrease  of $f_{\rm tr|s}$ with increasing
stellar  mass may also  simply reflect  that at  the massive  end most
central galaxies are already on  the red sequence.  We can distinguish
between   these  two  different   interpretations  by   examining  the
transition fraction of satellite galaxies  that were still blue at the
time of accretion.  This transition fraction is given by
\begin{equation}\label{ftrbluesat}
f_{\rm tr|s,bc}(\Ms) \equiv {f_{\rm tr|s}(\Ms) \over
                             f_{\rm b|c}(\Ms)} 
\end{equation} 
and  plotted  as  a  dashed  line  in the  lower  left-hand  panel  of
Fig.~\ref{fig:transit}.   Over  the  entire  range in  stellar  masses
probed,  this transition  fraction is  remarkably constant  at $f_{\rm
  tr|s,bc}  \simeq  0.4  \pm  0.1$.  Thus,  {\it  satellite  quenching
  affects roughly  40 percent of all  galaxies that are  still blue at
  their time of accretion, roughly independent of their stellar mass}.
\begin{table*}
\label{tab:fractions}
\caption{Fractions}
\begin{tabular}{rccccccccc}
   \hline
$\log(\Ms)$ & $f_{\rm s}$ & $f_{\rm s|r}$  & $f_{\rm s|b}$ & $f_{\rm r|c}$ & $f_{\rm r|s}$ & 
$f_{\rm tr|s}$ & $f_{\rm tr|s,bc}$ & $f_{\rm tr|r}$ & $f_{\rm tr|rs}$ \\
(1) & (2) & (3) & (4) & (5) & (6) & (7) & (8) & (9) & (10) \\ 
\hline\hline
 8.85 &  $31.4 \pm 1.1$ &  $58.4 \pm 3.7$ &  $24.6 \pm 1.1$ &  $12.2 \pm 1.4$ &  $37.4 \pm 3.2$ &  $25.2 \pm 3.4$ &  $28.7 \pm 3.7$ &  $39.4 \pm 5.0$ &  $67.5 \pm 4.5$ \\
 8.95 &  $33.0 \pm 1.3$ &  $61.8 \pm 3.4$ &  $24.7 \pm 0.7$ &  $12.7 \pm 1.0$ &  $41.8 \pm 2.6$ &  $29.1 \pm 2.8$ &  $33.3 \pm 3.1$ &  $43.0 \pm 4.1$ &  $69.6 \pm 3.0$ \\
 9.05 &  $34.2 \pm 1.2$ &  $64.1 \pm 3.1$ &  $24.2 \pm 1.0$ &  $13.7 \pm 1.3$ &  $47.1 \pm 2.2$ &  $33.4 \pm 2.6$ &  $38.7 \pm 2.8$ &  $45.5 \pm 4.1$ &  $70.9 \pm 3.1$ \\
 9.15 &  $33.5 \pm 1.0$ &  $61.5 \pm 3.2$ &  $23.2 \pm 0.8$ &  $15.7 \pm 1.5$ &  $49.5 \pm 2.0$ &  $33.8 \pm 2.7$ &  $40.1 \pm 2.8$ &  $42.0 \pm 4.3$ &  $68.3 \pm 3.5$ \\
 9.25 &  $33.4 \pm 1.1$ &  $60.7 \pm 2.3$ &  $22.7 \pm 1.1$ &  $16.7 \pm 1.2$ &  $51.3 \pm 1.9$ &  $34.7 \pm 2.4$ &  $41.6 \pm 2.5$ &  $41.0 \pm 3.1$ &  $67.6 \pm 2.7$ \\
 9.35 &  $31.9 \pm 0.8$ &  $52.1 \pm 1.8$ &  $23.0 \pm 0.7$ &  $21.5 \pm 1.0$ &  $49.9 \pm 1.4$ &  $28.4 \pm 1.7$ &  $36.2 \pm 2.0$ &  $29.6 \pm 2.2$ &  $56.9 \pm 2.4$ \\
 9.45 &  $32.5 \pm 0.7$ &  $53.9 \pm 1.6$ &  $22.2 \pm 0.6$ &  $22.3 \pm 1.0$ &  $54.1 \pm 1.4$ &  $31.7 \pm 1.6$ &  $40.9 \pm 1.9$ &  $31.6 \pm 1.9$ &  $58.7 \pm 2.0$ \\
 9.55 &  $31.5 \pm 0.7$ &  $50.1 \pm 1.7$ &  $21.3 \pm 0.6$ &  $25.7 \pm 0.9$ &  $56.2 \pm 1.4$ &  $30.5 \pm 1.8$ &  $41.0 \pm 2.1$ &  $27.2 \pm 1.9$ &  $54.3 \pm 2.2$ \\
 9.65 &  $30.1 \pm 0.5$ &  $47.1 \pm 0.9$ &  $19.8 \pm 0.6$ &  $28.7 \pm 0.7$ &  $59.1 \pm 0.8$ &  $30.4 \pm 1.2$ &  $42.6 \pm 1.4$ &  $24.2 \pm 1.1$ &  $51.5 \pm 1.5$ \\
 9.75 &  $29.7 \pm 0.5$ &  $44.6 \pm 0.9$ &  $19.1 \pm 0.6$ &  $32.7 \pm 0.8$ &  $62.3 \pm 1.2$ &  $29.6 \pm 1.4$ &  $44.0 \pm 1.9$ &  $21.2 \pm 1.1$ &  $47.5 \pm 1.6$ \\
 9.85 &  $29.1 \pm 0.6$ &  $41.0 \pm 1.0$ &  $19.0 \pm 0.4$ &  $38.1 \pm 0.4$ &  $64.5 \pm 0.9$ &  $26.5 \pm 1.0$ &  $42.7 \pm 1.5$ &  $16.8 \pm 0.8$ &  $41.0 \pm 1.1$ \\
 9.95 &  $28.5 \pm 0.4$ &  $39.1 \pm 0.7$ &  $17.6 \pm 0.4$ &  $43.3 \pm 0.6$ &  $69.7 \pm 0.7$ &  $26.4 \pm 1.0$ &  $46.5 \pm 1.4$ &  $14.8 \pm 0.6$ &  $37.9 \pm 1.1$ \\
10.05 &  $26.9 \pm 0.4$ &  $35.1 \pm 0.6$ &  $16.3 \pm 0.3$ &  $49.7 \pm 0.6$ &  $73.3 \pm 0.6$ &  $23.6 \pm 0.7$ &  $47.0 \pm 1.1$ &  $11.3 \pm 0.5$ &  $32.2 \pm 0.8$ \\
10.15 &  $26.5 \pm 0.4$ &  $33.8 \pm 0.6$ &  $15.5 \pm 0.3$ &  $54.1 \pm 0.3$ &  $76.7 \pm 0.5$ &  $22.5 \pm 0.7$ &  $49.2 \pm 1.2$ &   $9.9 \pm 0.4$ &  $29.4 \pm 0.7$ \\
10.25 &  $24.7 \pm 0.3$ &  $30.3 \pm 0.5$ &  $14.9 \pm 0.3$ &  $58.3 \pm 0.3$ &  $77.7 \pm 0.5$ &  $19.3 \pm 0.6$ &  $46.4 \pm 1.3$ &   $7.5 \pm 0.3$ &  $24.9 \pm 0.7$ \\
10.35 &  $22.9 \pm 0.2$ &  $27.5 \pm 0.4$ &  $14.3 \pm 0.2$ &  $61.2 \pm 0.2$ &  $78.1 \pm 0.4$ &  $16.9 \pm 0.5$ &  $43.6 \pm 1.3$ &   $6.0 \pm 0.2$ &  $21.7 \pm 0.6$ \\
10.45 &  $20.9 \pm 0.3$ &  $24.5 \pm 0.4$ &  $13.1 \pm 0.3$ &  $64.9 \pm 0.3$ &  $79.9 \pm 0.5$ &  $15.0 \pm 0.6$ &  $42.7 \pm 1.5$ &   $4.6 \pm 0.2$ &  $18.8 \pm 0.7$ \\
10.55 &  $19.2 \pm 0.3$ &  $22.2 \pm 0.4$ &  $12.0 \pm 0.3$ &  $68.0 \pm 0.3$ &  $81.7 \pm 0.5$ &  $13.7 \pm 0.5$ &  $42.9 \pm 1.5$ &   $3.7 \pm 0.2$ &  $16.8 \pm 0.6$ \\
10.65 &  $17.5 \pm 0.2$ &  $19.5 \pm 0.2$ &  $11.7 \pm 0.4$ &  $72.0 \pm 0.2$ &  $82.4 \pm 0.5$ &  $10.4 \pm 0.5$ &  $37.1 \pm 1.9$ &   $2.5 \pm 0.1$ &  $12.6 \pm 0.6$ \\
10.75 &  $14.9 \pm 0.2$ &  $16.1 \pm 0.2$ &  $10.4 \pm 0.3$ &  $77.3 \pm 0.2$ &  $84.9 \pm 0.4$ &   $7.6 \pm 0.4$ &  $33.5 \pm 1.9$ &   $1.4 \pm 0.1$ &   $9.0 \pm 0.5$ \\
10.85 &  $13.6 \pm 0.2$ &  $14.2 \pm 0.3$ &  $10.0 \pm 0.4$ &  $83.0 \pm 0.2$ &  $88.0 \pm 0.6$ &   $5.0 \pm 0.7$ &  $29.3 \pm 3.8$ &   $0.8 \pm 0.1$ &   $5.7 \pm 0.7$ \\
10.95 &  $11.1 \pm 0.3$ &  $11.4 \pm 0.4$ &   $8.7 \pm 0.6$ &  $88.8 \pm 0.2$ &  $91.4 \pm 0.7$ &   $2.6 \pm 0.8$ &  $23.1 \pm 6.7$ &   $0.3 \pm 0.1$ &   $2.8 \pm 0.8$ \\
11.05 &   $9.7 \pm 0.3$ &   $9.9 \pm 0.3$ &   $7.0 \pm 0.7$ &  $92.9 \pm 0.2$ &  $95.0 \pm 0.6$ &   $2.1 \pm 0.6$ &  $29.4 \pm 8.6$ &   $0.2 \pm 0.1$ &   $2.2 \pm 0.7$ \\
11.15 &   $7.6 \pm 0.3$ &   $7.7 \pm 0.3$ &   $5.9 \pm 0.8$ &  $94.6 \pm 0.2$ &  $95.9 \pm 0.7$ &   $1.3 \pm 0.6$ &  $24.2 \pm11.7$ &   $0.1 \pm 0.1$ &   $1.4 \pm 0.7$ \\
11.25 &   $5.7 \pm 0.3$ &   $5.8 \pm 0.3$ &   $3.2 \pm 1.9$ &  $96.1 \pm 0.2$ &  $97.8 \pm 1.3$ &   $1.7 \pm 1.3$ &  $44.5 \pm32.4$ &   $0.1 \pm 0.1$ &   $1.8 \pm 1.3$ \\
11.35 &   $3.5 \pm 0.4$ &   $3.5 \pm 0.4$ &   $1.7 \pm 1.8$ &  $97.2 \pm 0.4$ &  $98.6 \pm 1.4$ &   $1.4 \pm 1.3$ &  $51.1 \pm49.4$ &   $0.1 \pm 0.0$ &   $1.5 \pm 1.2$ \\
11.45 &   $2.1 \pm 0.6$ &   $2.1 \pm 0.7$ &   $0.0 \pm 0.0$ &  $97.5 \pm 0.6$ & $100.0 \pm 0.0$ &   $2.5 \pm 0.6$ & $100.0 \pm 0.0$ &  $0.1 \pm 0.0$ &   $2.5 \pm 0.6$ \\
\hline
\end{tabular}
\medskip

\begin{minipage}{\hdsize}
  {\it Notes:}  Various (transition)  fractions discussed in  the text
  and shown in Fig.~\ref{fig:transit}.   Column (1) lists the 10-based
  logarithm of the stellar mass (in $h^{-2} \Msun$), while columns~(2)
  to (10) list the various fractions plus their jackknife errors (both
  multiplied by 100).
\end{minipage}

\end{table*}

To express the overall impact  of satellite quenching for the build-up
of  the red  sequence,  we now  define  the fraction  of red  sequence
galaxies that arrived on the red sequence as a satellite galaxy. It is
straightforward to show that this fraction is given by
\begin{equation}\label{ftrredseq}
f_{\rm tr|r}(\Ms) \equiv {f_{\rm s}(\Ms) f_{\rm tr|s}(\Ms) \over
                          f_{\rm r}(\Ms)} 
\end{equation} 
and  is  shown as  a  solid  line in  the   lower right-hand  panel of
Fig.~\ref{fig:transit}.  It rapidly drops from   $\sim 40$ percent  at
$\Ms  = 10^9  h^{-2}  \Msun$ to $\sim  10$  percent at  $\Ms = 10^{10}
h^{-2} \Msun$ to   zero for $\Ms \gta   10^{11}  h^{-2} \Msun$.   This
clearly demonstrates   that  the vast  majority   of all red  sequence
galaxies have undergone their blue sequence to red sequence transition
as central  galaxies; even at a stellar  mass  of $10^9 h^{-2} \Msun$,
less than half of the galaxies on the red  sequence were quenched as a
satellite  galaxy. However,  part of the  reason  why this  transition
fraction  is so low  is simply  that the satellite  fraction  of (red)
galaxies is relatively low. Therefore, we  finally define the fraction
of red sequence  {\it satellite} galaxies that  has undergone the blue
sequence to red sequence transition  as a satellite. This fraction  is
given by
\begin{equation}\label{ftrredseqsat}
f_{\rm tr|rs}(\Ms) \equiv {f_{\rm s}(\Ms) f_{\rm tr|s}(\Ms) \over
                           f_{\rm rs}(\Ms)} = 
{f_{\rm tr|s}(\Ms) \over f_{\rm r|s}(\Ms)}
\end{equation} 
and  shown  as  the dashed  line  in  the  lower right-hand  panel  of
Fig.~\ref{fig:transit}.   Obviously  this   fraction  is  higher  than
$f_{\rm tr|r}$, reaching $\sim 70$ percent at $\Ms=10^9 h^{-2} \Msun$.
Therefore, at this relatively low  mass, only $\sim 30$ percent of the
satellite  galaxies on  the  red sequence  had  already been  quenched
before they  became a  satellite.  At the  massive end,  however, this
increases to virtually $100$ percent.

We have verified that all four transition fractions defined above only
depend very weakly on exactly how  we split our sample in red and blue
galaxies: increasing  or decreasing  $\calA_{\rm cut}$ by  0.02 yields
transition  fractions  that are  almost  indistinguishable from  those
shown in Fig.~\ref{fig:transit} within the errors.

Finally, as  a consistency check, we compare  the transition fractions
derived   here  with   the  average   color  differences   of  matched
central-satellite  pairs   presented  in  \S\ref{sec:transform}.   The
average  color  difference  is  related  to  the  transition  fraction
according to
\begin{eqnarray}  
\label{transit}   
\left\langle\grone_{\rm sat}-\grone_{\rm  cen} \right\rangle
& \approx & f_{\rm tr|s} 
\left(\calA_{\rm red}-\calA_{\rm blue}\right)\nonumber \\
& \simeq & 0.32 f_{\rm tr|s}
\end{eqnarray} 
Comparing the lower left-hand panel of Fig.~\ref{fig:transit} with the
upper panel   in the middle  column of  Fig.~\ref{fig:res} one can see
that $f_{\rm tr|s}(\Ms)$ and $\left\langle\grone_{\rm sat}-\grone_{\rm
    cen}   \right\rangle(\Ms)$ are     in   excellent  agreement  with
(\ref{transit}). For example, at  $\Ms \sim 3\times 10^9 h^{-2} \Msun$
the transition  fraction of  satellites is  $\sim 30$ percent,  which,
according   to  (\ref{transit}), corresponds   to   an  average  color
difference of $\sim 0.1$  magnitudes, in excellent agreement  with the
results shown in Fig.~\ref{fig:res}.

\subsection{Caveats}
\label{sec:caveats}

There  are a number  of caveats  with the  above determination  of the
relative importance of satellite  quenching.  The main assumption that
we have  made is that the  present day population  of central galaxies
can be considered representative of the progenitors of the present day
satellite galaxies.   However, in reality we should  have compared the
satellite  galaxies  to the  population  of  central  galaxies at  the
average redshift at which  the satellites were accreted.  Observations
have shown that  the mass density of galaxies on  the red sequence has
roughly doubled since  $z\sim 1$ (see \S~\ref{sec:intro}).  Therefore,
the red  fraction of centrals  is probably lower at  higher redshifts,
which  means  that  our  estimates  of the  transition  fractions  are
underestimated.  However, we don't believe  that this is a big effect. 
First of all, we have  focused on relatively massive satellites, which
most  likely have  only been  accreted relatively  recently, otherwise
they  would have  merged  with  the central  galaxy  due to  dynamical
friction.   This is  in  accord with  the  fact that  the present  day
population  of massive  subhaloes  (which host  the massive  satellite
galaxies)  fell into  their parent  halo fairly  recently  (Zentner \&
Bullock 2003; Gao \etal 2004; van  den Bosch, Tormen \& Giocoli 2005). 
Secondly,  in Appendix~A  we  use volume-limited  subsamples that  are
complete  in  stellar  mass  to  show that  there  is  no  significant
indication for any  evolution in the red fraction  of central galaxies
over the redshift range covered here ($0.01 \leq z \leq 0.2$).

Another questionable assumption that we have (implicitly) made is that
the  galaxy preserves  its stellar  mass once  it becomes  a satellite
galaxy.  In reality, it may continue  to form new stars (at least when
it remains  on the blue sequence) or  it may loose stars  due to tidal
stripping.  Since the latter is  likely to dominate, one would need to
compare the red fraction of satellites to the red fraction of centrals
that are more  massive, resulting in a lower  transition fraction.  To
have  some estimate  of  the  magnitude of  the  effect of  stripping,
consider  the case  in which  satellites have,  on average,  lost half
their stellar mass since  accretion. The transition fractions can then
be computed comparing  the red fraction of satellites  at stellar mass
$\Ms$  to the  red  fraction of  centrals  at $2  \Ms$. The  resulting
transition fractions are $\sim 10$ percent  lower at the low mass end. 
At the  massive end,  however, this results  in red fractions  for the
satellites  that   are  {\it  lower}  than  those   of  their  central
progenitors, resulting in negative transition fractions.  We therefore
argue that an  average mass loss rate of 50  percent is a conservative
upper  limit  and conclude  that  ignoring  mass  loss results  in  an
overestimate of  the various transition  fractions of no more  than 10
percent.

Finally,  there  is  a  potential  problem concerning  the  impact  of
interlopers, which  are galaxies assigned  to a group that  in reality
reside in a  different dark matter halo than the  other group members. 
Although our group  finding algorithm has been optimized  to yield low
interloper fractions  (see Yang \etal  2005a and Y07),  detailed tests
with mock  redshift surveys suggest  that we still have  an interloper
fraction  of roughly  $15\pm  5$  percent (see  Fig.~2  in Yang  \etal
2005a).   Most of  these interlopers  are  blue centrals  in low  mass
haloes  which  have been  erroneously  identified  as  a group  member
(satellite) of a bigger halo along  the line of sight to the observer. 
Thus,  interlopers tend  to  result  in an  overestimate  of the  blue
fraction of satellites, and hence  in an underestimate of the relative
importance  of satellite  quenching.   Since not  all interlopers  are
blue, though, we  estimate that the presence of  interlopers may cause
us to  underestimate the transition  fractions, $f_{\rm tr|s}$,  by no
more than 10 percent.

To summarize,  we have argued  that the three effects  mentioned here,
evolution, mass loss and interlopers, all have only a modest effect on
the inferred  transition fractions.  Furthermore,  since ignoring mass
loss  results   in  an  overestimate  while   ignoring  evolution  and
interlopers  results in  an  underestimate, the  cumulative effect  is
likely  to  be  small.   Nevertheless,  it  is  clear  that  potential
systematic  errors  can  be   significantly  larger  than  the  random
(jackknife) errors adopted here.  We intend to address these issues in
a  forthcoming paper  using large  mock  redshift surveys  based on  a
semi-analytical model for galaxy formation.

\section{Conclusions}
\label{sec:concl}

We  have  used  the SDSS  group  catalogue  of  Yang \etal  (2007)  to
investigate  the differences  between centrals  and satellites  of the
same stellar mass. Under the  hypothesis that a satellite galaxy was a
central galaxy  of the same stellar  mass before it  was accreted into
its  new host  halo, this  sheds light  on the  impact of  the various
transformation mechanisms  that are  believed to operate  on satellite
galaxies. Our conclusions can be summarized as follows:
\begin{itemize}
  
\item On average, satellite galaxies reside in dark matter haloes that
  are roughly twenty  times more massive than central  galaxies of the
  same stellar mass.
  
\item  On average, satellites  are somewhat  redder and  slightly more
  concentrated than  central galaxies of the same  stellar mass. There
  is a clear stellar mass dependence,  in the sense that the color and
  concentration    differences   are    larger   for    less   massive
  central-satellite pairs. In fact,  satellites with $\Ms \gta 10^{11}
  h^{-2} \Msun$ have average colors and concentrations that are almost
  identical to  those of  central galaxies of  the same stellar  mass. 
  This does  not imply, though, that the  transformation processes are
  less efficient for more massive galaxies. Rather, it simply reflects
  that virtually all galaxies with $\Ms \gta 10^{11} h^{-2} \Msun$ are
  already  red at  the time  they  are accreted  into a  larger halo.  
  Massive galaxies, therefore, must have undergone their late-to-early
  type transition while they were  still a central galaxy, most likely
  due to the impact of a major merger.
  
\item Central-satellite  pairs that are  matched in both  stellar mass
  and color show no  differences in concentration. However, pairs that
  are matched in both  stellar mass and concentration show significant
  color differences,  at least  for $\Ms \lta  10^{11} h^{-2}  \Msun$. 
  This  suggests  that  strangulation and/or  ram-pressure  stripping,
  which affect the  color, but not the concentration,  of a galaxy are
  more  efficient than  transformation mechanisms  that have  a strong
  impact on the galaxy's morphology, such as harassment.
  
\item   The   average   color   and   concentration   differences   of
  central-satellite  pairs  that  are  matched  in  stellar  mass  are
  independent of the halo mass  of the satellite.  This indicates that
  the  transformation  mechanism(s)  that  operate  on  the  satellite
  galaxies are equally efficient in haloes of all masses. This in turn
  argues against transformation mechanisms that are thought to operate
  only  in very  massive  haloes, such  as  ram-pressure stripping  or
  harassment.  Although these  processes do  occur, they  are  not the
  dominant cause of satellite quenching.
  
\item Comparing the  blue fractions of centrals and  satellites of the
  same  stellar mass  we infer  that roughly  40 percent  of  the blue
  galaxies transit to the red sequence after having been accreted into
  a bigger halo (i.e., after having become a satellite galaxy).  Since
  more massive  galaxies are less likely  to be blue at  the time they
  are accreted, the fraction of  {\it all} satellites that undergoes a
  transition  (between  their  time  of  accretion  and  the  present)
  decreases from  $\sim 35$  percent at $\Ms  = 10^9 h^{-2}  \Msun$ to
  basically zero percent at $\Ms = 10^{11} h^{-2} \Msun$.
  
\item Roughly 70 percent of satellite galaxies with $\Ms = 10^9 h^{-2}
  \Msun$ that  are on the red  sequence at the  present have undergone
  satellite quenching.   The remaining 30 percent were  already red at
  their time of accretion (i.e., when they became a satellite galaxy).
  For more  massive satellites, this relative  importance of satellite
  quenching is much lower. For  example, only $\sim 35$ percent of the
  red sequence  satellite galaxies with  $\Ms = 10^{10}  h^{-2} \Msun$
  have experienced  satellite quenching, while $\sim  65$ percent were
  already red at accretion. At  $\Ms = 10^{11} h^{-2} \Msun$ virtually
  all satellites were already red at accretion.

\end{itemize}
These   results  are  most   consistent  with   a  picture   in  which
strangulation  is  the  main  mechanism  that  operates  on  satellite
galaxies, and  that causes their transition  from the blue  to the red
sequence.   This conclusion  is consistent  (i) with  indications that
star formation  quenching takes place  on relatively long  time scales
(Kauffmann \etal 2004),  (ii) with the dependence of  the blue and red
fractions on  halo mass and galaxy luminosity  (Weinmann \etal 2006a),
(iii) with  recent simulation results  (Kawata \& Mulchaey  2007), and
(iv) with  a number of recent  studies that found  that star formation
quenching cannot be ascribed  solely to processes that are significant
only  in rich  clusters,  such as  ram-pressure  stripping and  galaxy
harassment (e.g.,  Balogh \etal 2002; Tanaka \etal  2004; Cooper \etal
2007;  Gerke  \etal  2007;  Verdugo,  Ziegler  \&  Gerken  2007).   

An   important,  outstanding   question   regards  the   morphological
transformations.   Since  most red  sequence  galaxies are  spheroids,
while  most  blue  sequence  galaxies  are disks,  it  is  clear  that
quenching alone  cannot explain the  build-up of the  red-sequence; an
additional  mechanism   is  required  to  transform   the  disks  into
spheroids. We  emphasize, therefore,  that strangulation is  the main,
not the  {\it only},  transformation mechanism operating  on satellite
galaxies.   Ram-pressure stripping,  harassment,  tidal stripping  and
heating, and  even satellite-satellite mergers  almost certainly occur
as well.   After all, as we  have shown, satellites  are slightly more
concentrated than  centrals of the same stellar  mass, indicating that
satellites  on average  undergo a  mild change  in  morphology.  Since
strangulation  only results  in  a quenching  of  the star  formation,
additional mechanisms  are required to explain  the full set  of data. 
If we make the simple  assumption that those galaxies that are already
red at the time of accretion are spheroids (since their transformation
mechanism is  most likely  related to a  major merger), while  the red
sequence galaxies that are  quenched as satellites maintain their disk
morphology, we  can use the results presented  in \S\ref{sec:trans} to
predict the spheroid fraction of  red sequence galaxies as function of
stellar mass:  it should  simply be $1-f_{\rm  tr|r}$, and  thus range
from unity for $\Ms \gta 10^{11} h^{-2} \Msun$ to $\sim 60$ percent at
$\Ms  = 10^9  h^{-2}  \Msun$.  Any  significant  deviations from  this
simple  prediction most likely  reflect the  impact of  the additional
satellite-specific transformation processes.  We intend to investigate
the morphological make-up of  the red-sequence (as function of stellar
mass) in a forthcoming study.

In most semi-analytical models  for galaxy formation (e.g., Kauffmann,
White \& Guiderdoni 1993; Somerville \& Primack 1999; Cole \etal 2000;
Croton \etal  2006; Kang \etal 2006) strangulation  is included (while
most  of the other  processes, such  as ram-pressure  stripping, tidal
heating and harassment are not).   However, as shown by Weinmann \etal
(2006b)  and Baldry  \etal  (2006), the  semi-analytical models  still
predict a red fraction for satellites  that is much too high (see Coil
\etal 2007 for a similar discrepancy at $z \sim 1$).  It is unclear at
the present  whether this  is due to  strangulation being  modeled too
efficiently, or whether  it reflects a problem with  the colors of the
infalling population.  In  all semi-analytical models strangulation is
modeled by instantaneously removing  the entire hot-gas reservoir of a
galaxy as soon as it  becomes a satellite.  However, recently McCarthy
\etal  (2007)   have  performed   a  large  suite   of  hydrodynamical
simulations and shown that $\sim  30\%$ of the hot gas associated with
a satellite galaxy remains bound to its subhalo for as long as 10 Gyr.
This  implies that the  satellites can  continue to  form stars  for a
longer  period (as  long as  the  hot gas  in the  subhalo can  cool),
resulting in  a less  efficient quenching of  the star  formation (see
also Kawata \&  Mulchaey 2007).  It remains to be  seen whether a more
realistic   description  of   strangulation,  consistent   with  these
simulation results, can successfully  fit the data presented here.  In
this  respect, including  similar constraints  from  higher redshifts,
such as those  provided by Cooper \etal (2006),  Cucciati \etal (2006)
and Gerke  \etal (2007),  will proof extremely  useful to  further our
understanding of the build-up of the red sequence.

\section*{Acknowledgments}

FvdB thanks the  Aspen Center for Physics where part  of this work has
been done, and is grateful to Eric Bell, Andreas Berlind, Alison Coil,
Charlie  Conroy, Hans-Walter  Rix, Erin  Sheldon, Jeremy  Tinker, Risa
Wechsler   and  Simon   White  for   enlightening   discussions.   DHM
acknowledges   support  from  the   National  Aeronautics   and  Space
Administration (NASA)  under LTSA Grant NAG5-13102  issued through the
Office of Space Science.



\appendix

\section[]{Redshift Evolution}
\label{sec:AppA}

Here  we illustrate  how  the  apparent magnitude  limit  of the  SDSS
results  in a stellar  mass limit  that depends  on both  redshift and
color, and  we describe the construction  of volume-limited subsamples
that  are complete  in stellar  mass,  and which  we use  to test  for
redshift evolution in the red fraction of centrals.
\begin{figure*}
\centerline{\psfig{figure=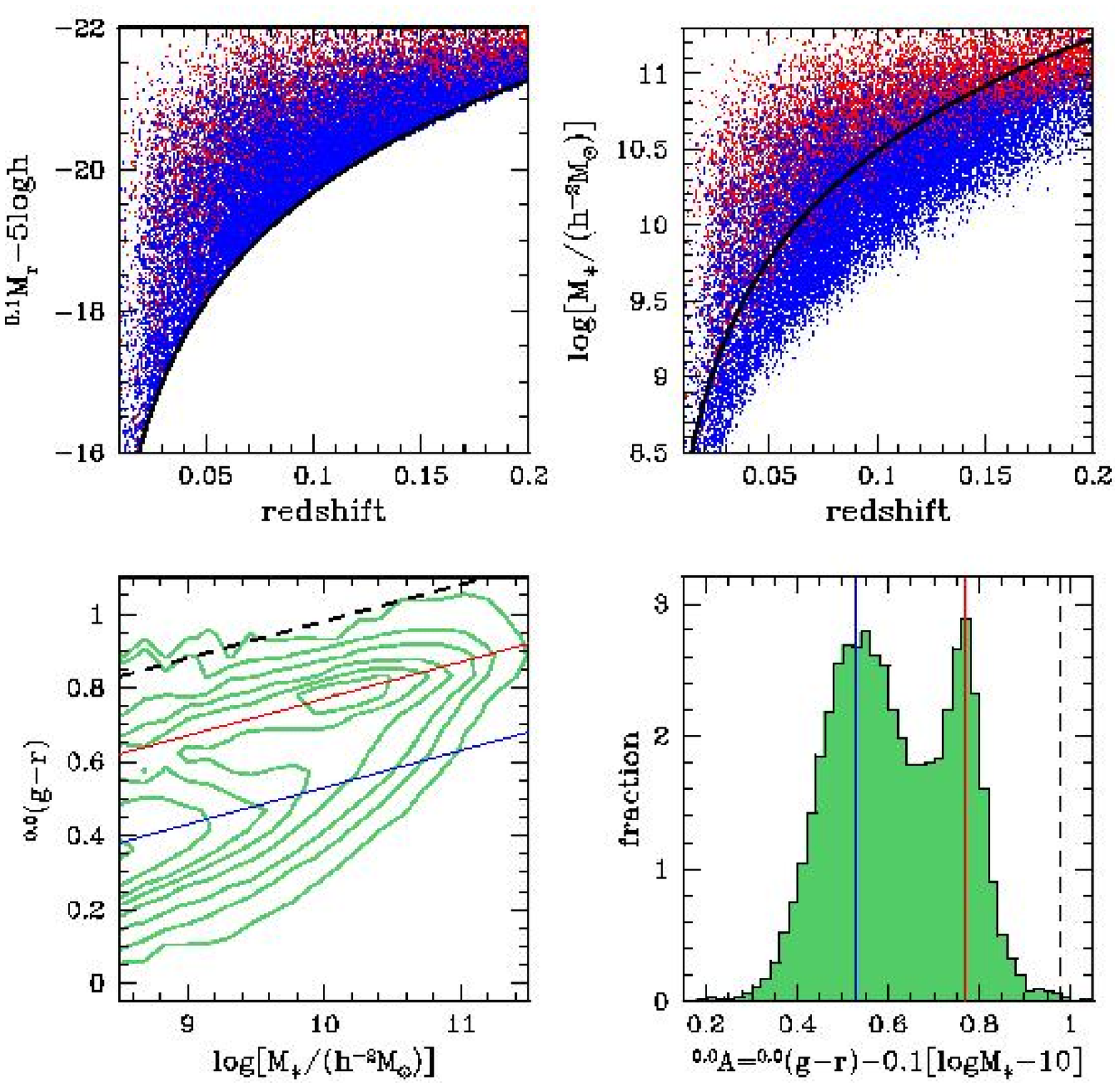,width=\hdsize}}
\caption{ {\it Upper left-hand panel:} The distribution of galaxies in
  sample~IIa  as function of  absolute ${^{0.1}r}$-band  magnitude and
  redshift (for clarity, only a  random subsample of 10 percent of all
  galaxies  is  shown).  Blue and  red  dots  refer  to blue  and  red
  galaxies,  based on  the criterion  discussed in  \S\ref{sec:trans}. 
  Note  that the $m_r=17.77$  apparent magnitude  limit of  the sample
  results  in a sharp,  redshift dependent  absolute magnitude  limit. 
  {\it  Upper right-hand panel:}  Same as  the upper  left-hand panel,
  except that  we now plot the  stellar mass as function  of redshift. 
  Note that  blue galaxies  (which have smaller  stellar mass-to-light
  ratios) can be seen out to higher redshifts than red galaxies of the
  same stellar mass.  The thick, black line indicates the stellar mass
  limit of eq.~(\ref{mstarlim}) above  which the sample is complete in
  stellar  mass.   {\it  Lower  left-hand panel:}  Contours  show  the
  $1/V_{\rm  max}$ weighted  color-stellar mass  distribution  of SDSS
  galaxies.   Contrary  to  Fig.~\ref{fig:cc_mstar}  we here  use  the
  $\grzero$ color,  which is used  to determine the stellar  masses of
  the galaxies  according to  eq.~(\ref{stelmass}).  The red  and blue
  solid lines indicate the red and blue sequences, respectively, while
  the  dashed  line  indicates  our  adopted  upper  envelope  of  the
  color-distribution.   {\it Lower  right-hand  panel:} The  $1/V_{\rm
    max}$ weighted distribution of $\calB$ for galaxies in our sample.
  The values of $\calB_{\rm  red}$, $\calB_{\rm blue}$ and $\calB_{\rm
    lim}$  are indicated  with red,  blue and  dashed  vertical lines,
  respectively.   Note the  pronounced  bimodality, and  that the  red
  sequence is significantly narrower than the blue sequence.}
\label{fig:mstarcut}
\end{figure*}

The  upper  left-hand   panel  of  Fig.~\ref{fig:mstarcut}  shows  the
absolute magnitude-redshift relation covering the redshift range $0.01
\leq z \leq 0.2$.  Galaxies  are color coded according to whether they
are red or  blue (based on the definition  of \S\ref{sec:trans}).  The
apparent magnitude limit of the sample ($m_r=17.77$) translates into a
redshift dependent absolute magnitude limit given by
\begin{eqnarray}
\label{magn01r}
\lefteqn{^{0.1}M_{r,{\rm lim}} - 5\log h =} \nonumber \\
&  & 17.77 - {\rm DM}(z) - k_{0.1}(z) + 1.62(z-0.1)\,.
\end{eqnarray}
where $k_{0.1}(z)$ is the $K$-correction to $z=0.1$, the $1.62(z-0.1)$
term is the evolution correction of Blanton \etal (2003), and
\begin{equation}
\label{distmeas}
{\rm DM}(z) = 5 \log D_L(z) + 25
\end{equation}
is the distance measure corresponding  to redshift $z$, with  $D_L(z)$
the    luminosity   distance   in  $h^{-1}\Mpc$.     As   discussed in
\S\ref{sec:data}, we  have used the   $K$-corrections of Blanton \etal
(2003; see also Blanton \& Roweis 2007), which  are obtained using the
five-band  photometry of  the   SDSS.    The redshift dependence    is
reasonably well described by
\begin{equation}
\label{kcorrect}
k_{0.1}(z) = 2.5\log\left({z+0.9 \over 1.1}\right)
\end{equation}
Substituting this in (\ref{magn01r}) yields the solid black line shown
in the upper left-hand  panel of Fig.~\ref{fig:mstarcut}.  At $z=0.1$,
where (by  definition) $k_{0.1} =  -2.5\log(1.1) \simeq -0.1$  for all
galaxies  (e.g.,  Blanton \&  Roweis  2007),  this  exactly gives  the
absolute  magnitude  limit  of   the  sample.   At  lower  and  higher
redshifts, however, a small fraction of the sample galaxies fall below
this limit.  This owes to  the fact that $k_{0.1}(z)$ not only depends
on  redshift but  also  on color.   For  the analyses  in this  paper,
however, this effect can safely be ignored.

As described in  Y07, stellar masses   for all galaxies in our  sample
have been computed using
\begin{eqnarray}
\label{stelmass}
\lefteqn{\log[\Ms/(h^{-2}\Msun)] = -0.406 + 1.097\left[\grzero\right]} 
\nonumber \\
& & - 0.4\left(^{0.0}M_r - 5\log h - 4.64\right)\,,
\end{eqnarray}
(see Bell  \etal 2003),  where $^{0.0}M_r -  5\log h$ is  the absolute
magnitude $K$-corrected and evolution corrected to $z=0.0$ using
\begin{equation}
\label{magnr}
^{0.0}M_r - 5\log h = m_r - {\rm DM}(z) - k_{0.0}(z) + 1.62 z\,.
\end{equation}
The  upper  right-hand  panel  of  Fig.~\ref{fig:mstarcut}  shows  the
resulting stellar mass  as function of redshift, using  the same color
coding as in  the upper left-hand panel.  Note that  there is no sharp
limit to the stellar mass distribution at any given redshift, and that
the low mass end is  entirely dominated by blue galaxies.  This arises
because blue  galaxies have a  lower stellar mass-to-light  ratio than
red  galaxies of  the same  stellar mass,  which is  expressed  by the
color-term in (\ref{stelmass}).   Therefore, in a flux-limited sample,
blue galaxies  can be seen out  to higher redshifts  than red galaxies
{\it  of  the  same  stellar  mass}.  In  the  analysis  described  in
\S\ref{sec:trans} this  bias has been  corrected for by  weighting all
galaxies by $1/V_{\rm max}$ when computing red and blue fractions.

To construct volume-limited  subsamples that  are complete in  stellar
mass, we proceed as follows. Since redder galaxies have higher stellar
mass-to-light ratios, we first determine a suitable upper limit to the
color distribution of galaxies as function  of their stellar mass. The
lower left-hand panel  of Fig.~\ref{fig:mstarcut} shows a contour plot
of  the  $1/V_{\rm max}$ weighted   distribution of   galaxies in  the
parameter  space   of  $\grzero$ versus   stellar    mass.  Similar to
Fig.~\ref{fig:cc_mstar},   it  clearly  reveals  the   bimodal   color
distribution of the galaxy   distribution.  Fitting the   stellar mass
dependence of the red and blue sequences (by eye) we obtain
\begin{equation}\label{collim}
\grzero = \calB + 0.10 \left( \log[\Ms/(h^{-2}\Msun)] - 
10.0\right)\,,
\end{equation}
with  $\calB=\calB_{\rm red}=0.77$  and $\calB=\calB_{\rm  blue}=0.54$
for  the  red and  blue   sequences, respectively.   The   dashed line
corresponds to  $\calB=\calB_{\rm  lim}=0.98$ and roughly  reflects the
upper  limit to  the color distribution   (only $0.3$ percent of  the
galaxies in  our sample  have  colors  redder than  this).   This  is
further   illustrated    in     the  lower    right-hand  panel     of
Fig.~\ref{fig:mstarcut}   which  plots  the $1/V_{\rm   max}$-weighted
distribution of $\calB = \grzero - 0.10 \left( \log[\Ms/(h^{-2}\Msun)]
  -  10.0\right)$ for  all galaxies   in  our sample.   Note that  the
bi-modality of the color  distribution is now  extremely clear, as is
the fact that the red sequence is significantly narrower than the blue
sequence.
\begin{figure}
\centerline{\psfig{figure=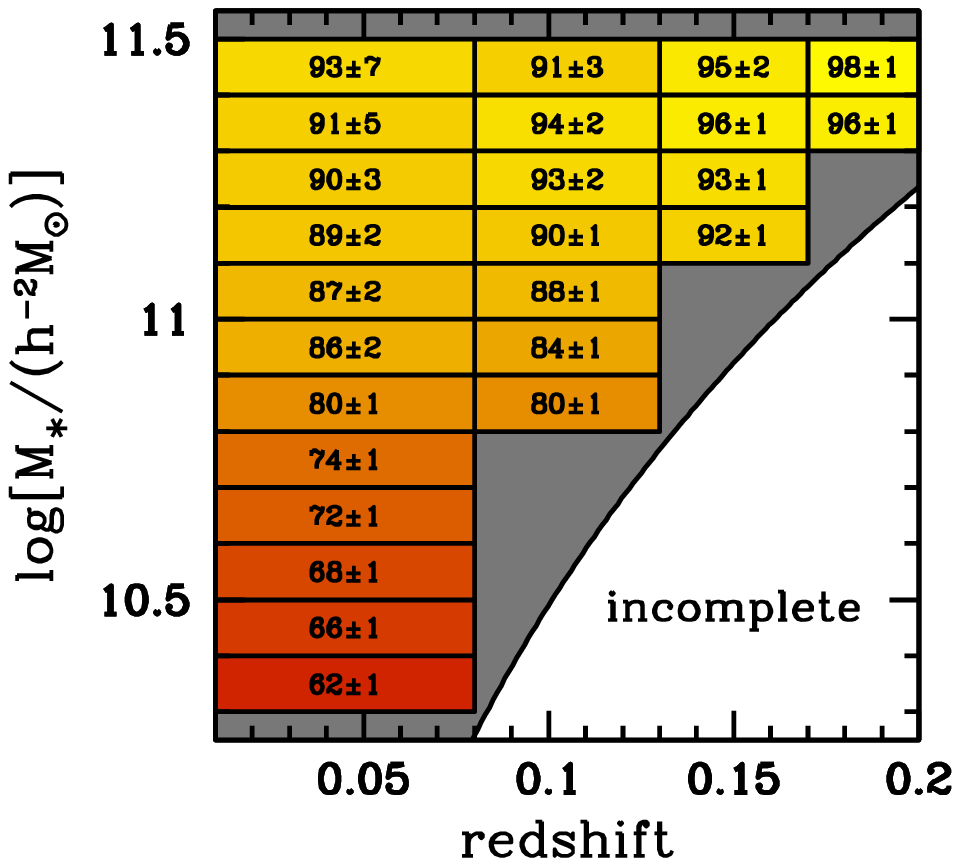,width=\hssize}}
\caption{Redshift and stellar mass dependence of the red fraction of
  centrals. The gray  area indicates the parameter space  where we are
  complete  in  stellar  mass  (i.e.,,  the  region  above  $M_{*,{\rm
      lim}}(z)$ given  by eq.~[\ref{mstarlim}]), and  the shaded boxes
  indicate  volume-limited  subsamples that  are  complete in  stellar
  mass.   They are  color coded  according  to their  red fraction  of
  centrals,  the value  and (Poisson)  error of  which are  indicated. 
  Note that there is no significant indication of redshift evolution.}
\label{fig:mlim}
\end{figure}

The final  ingredient required for the construction  of volume limited
sub-samples  that are  complete in  stellar mass  is a  simple fitting
function for $k_{0.0}(z)$.  After some experimenting we found that the
color and redshift dependence  of $k_{0.0}(z)$ can be reasonably well
described by
\begin{equation}
\label{k00corr}
k_{0.0}(z) = 2.5\log(1.0+z) + 1.5 z \left[\grzero - 0.66\right]
\end{equation}

Combining   (\ref{stelmass})-(\ref{k00corr})  and adopting $m_r=17.77$
and   $\calB=0.98$, we  obtain a relation   between   stellar mass and
redshift given by
\begin{eqnarray}
\label{mstarlim}
\lefteqn{\log[M_{*,{\rm lim}}/(h^{-2}\Msun)] =} \\
 & & {4.852 + 2.246 \log D_L(z) + 1.123 \log(1+z) - 1.186 z \over 1 - 0.067
  z} \nonumber
\end{eqnarray}
and indicated by  the thick, solid line in  the upper right-hand panel
of   Fig.~\ref{fig:mstarcut}.  We   can   now  define   volume-limited
subsamples  that  are complete  in  stellar  mass,  by only  selecting
galaxies   that  lie   above  this   limit.   The   shaded   boxes  in
Fig.~\ref{fig:mlim} indicate such  samples.  The boxes are color-coded
according to their red fraction  of centrals, $f_{\rm r|c}$, the value
and  (Poisson) error  of which  are  indicated. Note  that within  the
errors there  is no indication for any  significant redshift evolution
in  $f_{\rm r|c}$,  though  it is  clear  that the  redshift-dependent
stellar  mass   limit  of  eq.~(\ref{mstarlim})  only   allows  us  to
investigate this for the most massive centrals.

\end{document}